\documentclass[preprint,12pt,times]{elsarticle}
\usepackage{amsmath,amssymb,xspace,subfigure}
\usepackage{amsfonts}
\usepackage[usenames,dvipsnames]{xcolor}
\usepackage[]{graphicx}
\usepackage{url}
\usepackage{hyperref}
\usepackage{fullpage}
\usepackage{url}
\usepackage{float}
\usepackage{bm}
\usepackage{soul}
\usepackage{natbib}
\usepackage{soul}
\usepackage{setspace}

\newcommand{\blue}[1]{{\color{blue} #1}}

\graphicspath{{./Figures/}}

\journal{Elsevier}

\begin{document}

\begin{frontmatter}

%% Title, authors and addresses

%% use the tnoteref command within \title for footnotes;
%% use the tnotetext command for the associated footnote;
%% use the fnref command within \author or \address for footnotes;
%% use the fntext command for the associated footnote;
%% use the corref command within \author for corresponding author footnotes;
%% use the cortext command for the associated footnote;
%% use the ead command for the email address,
%% and the form \ead[url] for the home page:
%%
%% \title{Title\tnoteref{label1}}
%% \tnotetext[label1]{}
%% \author{Name\corref{cor1}\fnref{label2}}
%% \ead{email address}
%% \ead[url]{home page}
%% \fntext[label2]{}
%% \cortext[cor1]{}
%% \address{Address\fnref{label3}}
%% \fntext[label3]{}

\title{Discrete differential geometry-based model for nonlinear analysis of axisymmetric shells}

%% use optional labels to link authors explicitly to addresses:
%% \author[label1,label2]{<author name>}
%% \address[label1]{<address>}
%% \address[label2]{<address>}

\author[a]{Weicheng Huang\fnref{label2}}
\author[b]{Tianzhen Liu\fnref{label2}}
\author[c]{Zhaowei Liu}
\author[d]{Peifei Xu}
\author[e,f]{Mingchao Liu\corref{cor5}}
\author[g]{Yuzhen Chen\corref{cor5}}
\author[f,h]{K. Jimmy Hsia\corref{cor5}}
\fntext[label2]{W.H. and T.L. contributed equally to this work.}
\cortext[cor5]{Corresponding authors: m.liu.2@bham.ac.uk (M.L.), yuzhenchen@fudan.edu.cn (Y.C), kjhsia@ntu.edu.sg (K.J.H)}

\address[a]{School of Engineering, Newcastle University, Newcastle upon Tyne NE1 7RU, UK}
\address[b]{Key Laboratory of C $\&$ PC Structures of Ministry of Education, National Prestress Engineering Research Center,  \\ Southeast University, Nanjing 210096, China}
\address[c]{Department of Engineering Mechanics, Hohai University, Nanjing 211000, China}
\address[d]{School of Mechanical Engineering, Southeast University, Nanjing 211189, China}
\address[e]{Department of Mechanical Engineering, University of Birmingham, Birmingham B15 2TT, UK}
\address[f]{School of Mechanical and Aerospace Engineering, Nanyang Technological University, \\ Singapore 639798, Republic of Singapore}
\address[g]{Institute of Mechanics and Computational Engineering, Department of Aeronautics and Astronautics, \\ Fudan University, Shanghai 200433, China}
\address[h]{School of Chemistry, Chemical Engineering and Biotechnology, Nanyang Technological University, \\ Singapore 639798, Republic of Singapore}

\begin{abstract}

In this paper, we propose a novel one-dimensional (1D) discrete differential geometry (DDG)-based numerical method for geometrically nonlinear mechanics analysis (e.g., buckling and snapping) of axisymmetric shell structures.
Our numerical model leverages differential geometry principles to accurately capture the complex nonlinear deformation patterns exhibited by axisymmetric shells.
By discretizing the axisymmetric shell into interconnected 1D elements along the meridional direction, the in-plane stretching and out-of-bending potentials are formulated based on the geometric principles of 1D nodes and edges under the Kirchhoff–Love hypothesis, and elastic force vector and associated Hession matrix required by equations of motion are later derived based on symbolic calculation.
Through extensive validation with available theoretical solutions and finite element method (FEM) simulations in literature, our model demonstrates high accuracy in predicting the nonlinear behavior of axisymmetric shells.
Importantly, compared to the classical theoretical model and three-dimensional (3D) FEM simulation, our model is highly computationally efficient, making it suitable for large-scale real-time simulations of nonlinear problems of shell structures such as instability and snap-through phenomena. 
Moreover, our framework can easily incorporate complex loading conditions, e.g., boundary nonlinear contact and multi-physics actuation, which play an essential role in the use of engineering applications, such as soft robots and flexible devices.
This study demonstrates that the simplicity and effectiveness of the 1D discrete differential geometry-based approach render it a powerful tool for engineers and researchers interested in nonlinear mechanics analysis of axisymmetric shells, with potential applications in various engineering fields.

\end{abstract}


\end{frontmatter}

\section{Introduction}
\label{intro}

Shell formations are widespread in both natural settings, such as biological vesicles, cells, and fruits, as well as in artificial designs like footballs, submarines, and space capsules.
Many of these structures exhibit axisymmetric characteristics, owing to their intrinsic symmetry in geometry and loading conditions \cite{nasir2002axisymmetric}.
Mechanical behaviors of axisymmetric shells hold paramount significance across diverse engineering applications, encompassing biomedical devices, pressure vessels, and aerospace structures.
Understanding the complex deformation and performance of such shells under different loading is essential to ensure their intended functionality and operational safety.
One tragic example of shell failure is the implosion of the deep-sea submersible Titan on June 30, 2023, resulting in a great loss of life and highlighting the critical importance of structural integrity, materials, and engineering constraints of axisymmetric shells \cite{bbc2023titan}.

One common approach for the mechanical analysis of axisymmetric shell systems is to utilize the classical shell theory \cite{hutchinson2017nonlinear,koiter1969nonlinear}, which can provide accurate predictions on the nonlinear responses, including the instability behaviors \cite{xirouchakis1980axisymmetric,yan2020buckling,hutchinson2016buckling,yang2018pattern,qiao2020elastic,liu2022buckling}.
However,  shell theories are mostly based on the small-strain, moderate rotation assumptions \cite{sanders1963nonlinear}. 
When it comes to shell snapping or everting, these assumptions no longer hold since large deflections and rotations occur as the shell turns inside-out.
Moreover, directly solving ordinary differential equations (ODEs) of the shell theory via functional variation is highly complicated or even impractical, especially when the structure is subjected to complex loading, geometric nonlinearity, nonlinear boundary conditions, or multi-physical interactions.
Consequently, using shell theories to directly solve nonlinear deformation problems of shells would be extremely challenging. 

On the other side, the Finite Element Method (FEM), which handles the large deformation and finite rotation by imposing the nonlinear geometry formulation, has been widely employed to study nonlinear behaviours of shells, such as buckling and snapping, under different loading conditions \cite{taffetani2018static, pezzulla2019weak,nasikas2022framework}, as well as design shell-based functional devices \cite{abdullah2016programmable,gorissen2020inflatable,hao2020progressive,qiao2021bi}.
Despite its versatility, the FEM suffers from high computational costs, and challenges if the deformation is triggered by complex loading conditions such as magnetic actuation \cite{yan2021magneto,stewart2023magneto,dadgar2023micropolar,abbasi2023leveraging}, dynamic loading \cite{gorissen2020inflatable}, and contact \cite{zhang2020tunable}.
The challenges of studying the shell-snapping phenomena using the above methods call for a highly efficient yet simple-to-implement numerical method.

A recent developed numerical framework – Discrete Differential Geometry (DDG) formulation, which is popular for picture animation in movie special effects due to its computational efficiency \cite{grinspun2006discrete} – is being adopted and extended for mechanical analysis of slender elastic structures.
DDG-based methods involve discretizing a smooth structure into a mass-spring-type system while preserving the crucial geometric and physical properties of the actual object.
As the local coordinate-independent geometric properties (e.g., strains and curvatures) are used to construct the elastic potentials of the deformable body, the geometrically nonlinear deformations such as large deflections and large rotations can be directly incorporated within this framework, resulting in the fully nonlinear numerical model.
Moreover, previous studies using the DDG-based numerical framework have been very successful at simulating slender elastic structures such as beams \cite{lestringant2020discrete}, rods \cite{bergou2008discrete,bergou2010discrete}, ribbons \cite{audoly2021one,charrondiere2020numerical}, gridshells \cite{panetta2019x}, plates \cite{savin2011growth}, and shells \cite{grinspun2003discrete}, as well as axisymmetric membranes \cite{turlier2014furrow}, demonstrating its capabilities for simulating engineering problems, e.g., cable deployment \cite{jawed2014coiling}, strip bifurcations \cite{huang2022discrete}, gridshell form finding \cite{baek2018form}, cell cytokinesis \cite{turlier2014furrow}, and soft robot dynamics \cite{huang2020dynamic}.

Here, we formulate a novel DDG-based numerical model of a thin, elastic, axisymmetric shell undergoing axisymmetric deformation.
Due to symmetry, an axisymmetric shell surface can be represented by a single curve rotating with respect to the central axis for a complete $360^{\circ}$ revolution.
Such a curve can be characterized by a curvilinear coordinate $s$, thus reducing it to a 1D problem \cite{taffetani2018static, pezzulla2019weak, turlier2014furrow}.
With the nodal-based description of the rotational curve, the DDG-based approach can automatically capture the fully geometric nonlinearity of the flexible structures, e.g., large deflections and rotations.
The discrete elastic potential is formulated under the classical Kirchhoff–Love hypothesis, and the elastic force vector and the associated Hession matrix required by the dynamic equations of motion are then derived analytically through symbolic calculation, resulting in a fully implicit algorithm with time complexity $O(N)$.
Several examples from existing literature are used for cross-validation.
Importantly, our numerical tool shows better accuracy when compared with classical shell equations, and can also be potentially faster than in real-time.
Moreover, this type of nodal-based discrete simulation is naturally suited to incorporate different loading conditions, e.g., combining incremental potential method for boundary nonlinear contact \cite{li2020incremental, huang2023modeling} and employing Helmholtz’s free energy theory to encapsulate the magnetic actuation \cite{yan2021magneto, pezzulla2022geometrically, huang2023discrete}, which are important for both scientific research and engineering applications.
We here use two examples to demonstrate that our numerical method can easily handle complex loading conditions and be applied to real engineering problems: (i) contact-induced shell snapping and jumping, (ii) magnetic-induced shell buckling and snapping.

This paper is organized as follows.
In Section 2, we introduce the discrete model for the numerical analysis of 1D axisymmetric shell structure undergoing nonlinear deformation.
Then, several benchmarks are used to validate our newly introduced framework in Section 3.
Next, in Section 4, our numerical tool is combined with boundary nonlinear contact and multi-physics actuation for the analysis of real engineering problem.
Finally, concluding remarks and avenues for future research are presented in Section 5.
We also review the classical 1D shell theory in {\color{blue}Appendix A}, perform a convergence study for the contact model in {\color{blue}Appendix B}, and provide the details of FEM simulations for validation in {\color{blue}Appendix C}.

\section{Numerical model}\label{sec:discreteModel}

Here, we will begin by providing a brief overview of the general Kirchhoff-Love shell model and its fundamental principles. We subsequently develop a 1D reduced-order, discrete model for spherical shells, specifically tailored to address the axisymmetric condition.
We further incorporate various loading conditions for a comprehensive analysis of the spherical shell's mechanical behaviors.

\begin{figure}[h!]
  \centering
  \includegraphics[width=0.6\textwidth]{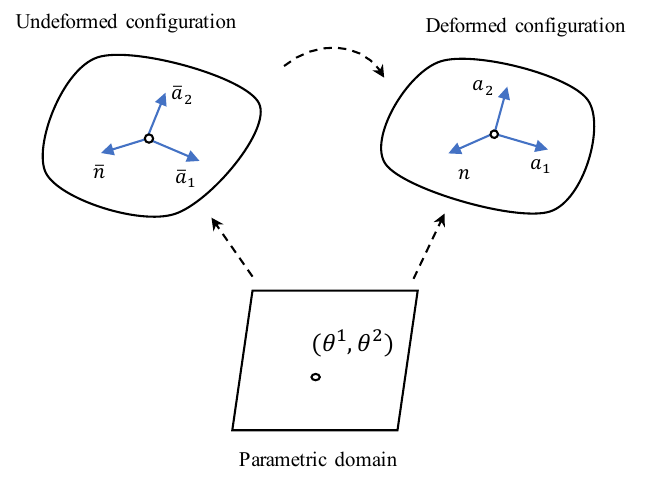}
  \caption{Mid-surfaces of the undeformed and the deformed shell element. Their covariant basis are $\{ \bar{\mathbf{a}}_1, \bar{\mathbf{a}}_2 \}$ and $\{ \mathbf{a}_1, \mathbf{a}_2\}$, respectively.}
  \label{fig:modelShellPlot}
\end{figure}

\subsection{General shell model}

A basic hypothesis of the geometrically nonlinear Kirchhoff-Love shell model \cite{pezzulla2019weak, guo2019isogeometric} is that the lines perpendicular to the mid-surface of a thin shell remain straight after deformation. As a result, one can describe both the undeformed and deformed configurations using a 2D surface $\omega = {\mathbf x} (\theta^1, \theta^2) \in \mathcal{R}^3$, parameterized by $(\theta^1, \theta^2)$, as illustrated in Fig.~\ref{fig:modelShellPlot}. Here the Greek indices ${\alpha, \beta}$ take values $\{1, 2\}$, following the convention.
Based on this assumption, a material point on the mid-surface in the undeformed and deformed configurations is denoted by $\bar{\mathbf x}$ and $\mathbf x$, respectively.
Hereafter, a bar on top of a symbol denotes that parameter in its undeformed configuration.
The mid-surface point in the deformed configuration ${\mathbf x}$ can be related to the mid-surface point in the reference configuration, as
\begin{equation}
\mathbf x = \bar{\mathbf x} + \mathbf u,
\label{eq:reference_to_deformed}
\end{equation}
where $\mathbf u$ is the displacement vector.
The covariant basis vectors for both the deformed and undeformed surfaces are
\begin{equation}
{\mathbf a}_{\alpha} = \frac{\partial {\mathbf x}}{ \partial \theta^{\alpha}}
\quad \text{and} \quad
{ \bar{\mathbf a}}_{\alpha} = \frac{\partial { \bar {\mathbf x}} }{ \partial \theta^{\alpha}},
\end{equation}
and the associated covariant metric coefficients can be determined as the dot product between the covariant basis vectors,
\begin{equation}
{\mathbf a}_{\alpha \beta} = \mathbf{a}_{\alpha} \cdot \mathbf{a}_{\beta}
\quad \text{and} \quad
\bar{{\mathbf a}}_{\alpha \beta} = \bar{\mathbf{a}}_{\alpha} \cdot \bar{\mathbf{a}}_{\beta}.
\end{equation}
The components of the contravariant matrix can be determined as its inverse, as,
\begin{equation}
{\mathbf a}^{\alpha \beta} = ({\mathbf a}_{\alpha \beta})^{-1}
\quad \text{and} \quad
\bar{\mathbf a}^{\alpha \beta} = ( \bar{\mathbf a}_{\alpha \beta})^{-1},
\end{equation}
such that the contravariant basis vectors can be
computed as
\begin{equation}
{\mathbf a}^{\alpha} = {\mathbf a}^{\alpha \beta} \cdot \mathbf{a}_{\beta}
\quad \text{and} \quad
\bar{\mathbf a}^{\alpha} = \bar{\mathbf a}^{\alpha \beta} \cdot \bar{\mathbf{a}}_{\beta}.
\end{equation}
The normal unit vector of the mid-surface is given by the cross-product between the two covariant basis vectors, as
\begin{equation}
{\mathbf n} = {\mathbf a}_3 = \frac{{\mathbf a}_1 \times {\mathbf a}_2}{|| {\mathbf a}_1 \times {\mathbf a}_2||} \equiv {\mathbf a}^3
\quad \text{and} \quad
\bar{\mathbf n} = \bar{\mathbf a}_3 = \frac{ \bar{\mathbf a}_1 \times \bar{\mathbf a}_2}{|| \bar{\mathbf a}_1 \times \bar{\mathbf a}_2||} \equiv \bar{\mathbf a}^3.
\end{equation}
The first and second fundamental forms of the mid-surface are updated based on the covariant basis vectors and the surface normal vector,
\begin{equation}
\begin{aligned}
A_{\alpha\beta} = \frac {1} {2} \left( \mathbf{a}_\alpha \cdot \mathbf{a}_\beta \right) 
\quad &\text{and} \quad
\bar{A}_{\alpha\beta} = \frac {1} {2} \left( \bar{\mathbf{a}}_\alpha \cdot \bar{\mathbf{a}}_\beta \right) 
\\
B_{\alpha\beta} = \mathbf{a}_{\alpha,\beta} \cdot \mathbf n
\quad &\text{and} \quad
\bar{B}_{\alpha\beta} = \bar{\mathbf{a}}_{\alpha,\beta} \cdot \bar{\mathbf n},
\end{aligned}
\end{equation}
where 
\begin{equation}
\mathbf{a}_{\alpha,\beta} = \frac {\partial \mathbf{a}_{\alpha} } {\partial \theta^{\beta}}
\quad \text{and} \quad
\bar{\mathbf{a}}_{\alpha,\beta} = \frac { \partial \bar{\mathbf{a}}_{\alpha} } {\partial \theta^{\beta}}.
\end{equation}
Next, the in-plane stretching strain tensor and the out-of-plane bending curvature tensors are given by the differences between the deformed and undeformed configurations,
\begin{equation}
\begin{aligned}
\epsilon_{\alpha \beta} &= A_{\alpha\beta} - \bar{A}_{\alpha\beta} \\
\kappa_{\alpha \beta} &= B_{\alpha\beta} - \bar{B}_{\alpha\beta}.
\end{aligned}
\end{equation}
Finally, for a shell with thickness $h$, of linear elastic material with Young's modulus $E$ and Poisson's ratio $\nu$, its total elastic potential takes the quadratic form of strains, as
\begin{equation}
U = \iint_{\omega} \; \frac{1}{2} \; \bm{\epsilon} \cdot \mathbb{D}_{s} \cdot \bm{\epsilon} \; d\omega + \iint_{\omega} \; \frac{1}{2} \; \bm{\kappa} \cdot \mathbb{D}_{b} \cdot \bm{\kappa} \; d\omega,
\end{equation}
where the stiffness tensors are
\begin{equation}
\mathbb{D}_{s} = \frac{Eh}{1-\nu^2}
\begin{bmatrix}
1 & \nu \\
\nu & 1
\end{bmatrix}
\quad \text{and} \quad
\mathbb{D}_{b} = \frac{Eh^3}{12(1-\nu^2)}
\begin{bmatrix}
1 & \nu \\
\nu & 1
\end{bmatrix}.
\end{equation}
Based on these definitions of the Kirchhoff-Love shell, governing equations can be derived using the variational approach and, by prescribing associated boundary conditions, the equilibrium solution can be eventually obtained.

\subsection{Reduced 1D discrete shell model}

\begin{figure}[t!]
  \centering
  \includegraphics[width=0.8\textwidth]{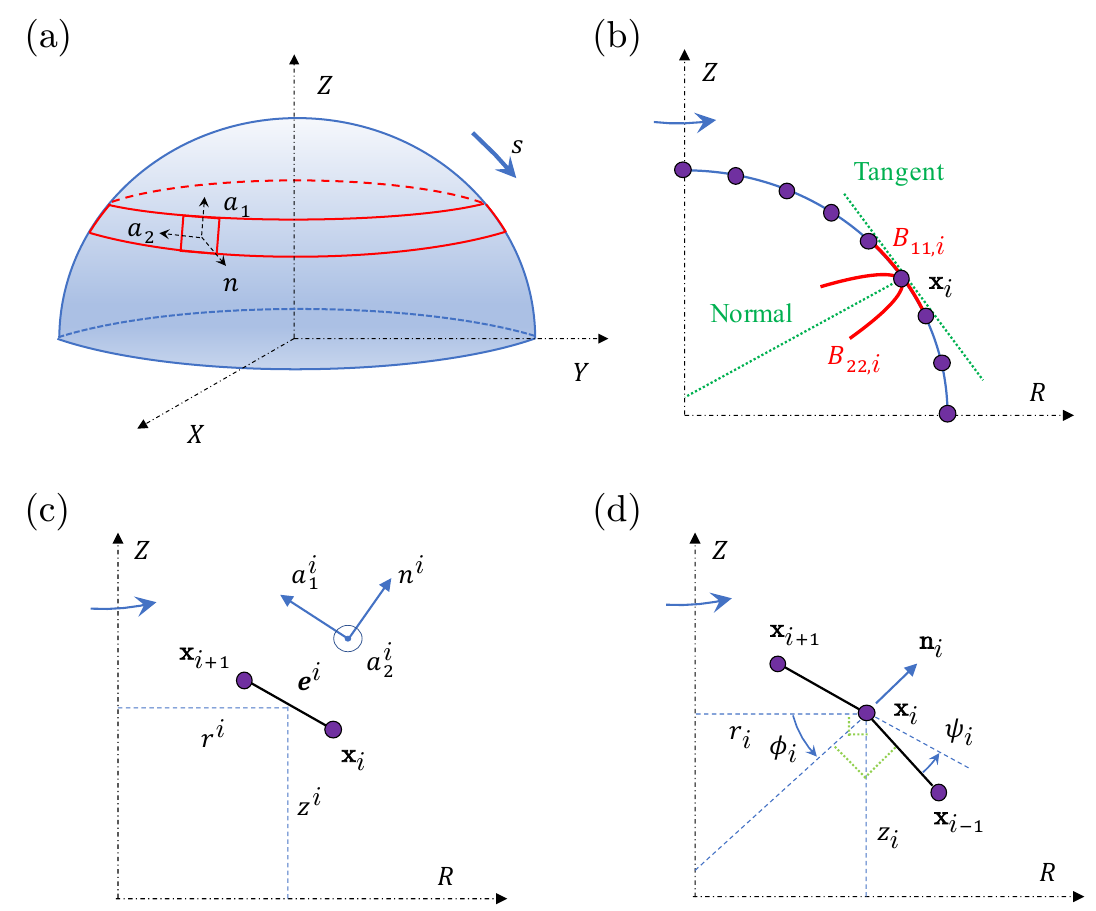}
  \caption{Discretization of axisymmetric shell using DDG-based method. (a) 3D configuration of an axisymmetric shell. (b) The corresponding 1D representation of the shell in the $R-Z$ plane. Note that its second fundamental form is marked by red. (c) Discrete format of $i$-th edge and its local frame. (d) Discrete curvature representation for $i$-th bending element.}
  \label{fig:modelPlot}
\end{figure}

We present the procedure for a discrete numerical model of axisymmetric shells based on the DDG framework \cite{grinspun2006discrete}.
An axisymmetric thin shell can be represented by a curvilinear coordinate $s = {\mathbf x} (r, z) \in \mathcal{R}^2$ (Fig.~\ref{fig:modelPlot}(a) and Fig.~\ref{fig:modelPlot}(b)).
We denote the meridional direction as the 1-direction, $\mathbf{a}_{1}$, the circumferential direction as the 2-direction, $\mathbf{a}_{2}$, and the surface normal vector is their cross-product.
It is clear that the meridional and the circumferential directions are not coupled due to symmetry.

\paragraph{Kinematics} We discretize the spherical shell contour along $s$ into $N$ consecutive nodes, giving rise to a vector $\mathbf{q} \in \mathbf{R}^{2N \times 1}$ in $R-Z$ space, as 
\begin{equation}
\mathbf{q} = [ \mathbf{x}_{0}, \mathbf{x}_{1}, ..., \mathbf{x}_{i}, ..., \mathbf{x}_{N-1} ],
\label{dofVector}
\end{equation}
where $\mathbf{x}_{i} \equiv [ r_{i}, z_{i} ]$.
The edge vectors connecting the nodes are defined by
\begin{equation}
\mathbf{e}^{i} = \mathbf{x}_{i+1} - \mathbf{x}_{i},
\end{equation}
and, as shown in Fig.~\ref{fig:modelPlot}(c), its tangent vector and normal vector satisfy
\begin{equation}
\mathbf{t}^{i} = \frac {\mathbf{e}^{i} } {l^{i}}
\quad \text{and} \quad
\mathbf{n}^{i} \cdot \mathbf{t}^{i} = 0.
\end{equation}
where $l^{i} = || \mathbf{e}^{i} ||$ is its length and is computed by the $\mathcal{L}_{2}$ norm of the edge vector.
Hereafter, we use subscripts to denote quantities associated with the nodes, e.g., $\mathbf{x}_{i}$, and superscripts when associated with edges, e.g., $\mathbf{e}^{i}$, and the average is used to transfer the node-based quantity to the edge-based quantity, e.g., in Fig.~\ref{fig:modelPlot}(c), the radius of $i$-th edge is denoted as 
\begin{equation}
r^{i} = \frac{1}{2} (r_{i}+r_{i+1}), 
\end{equation}
and the Voronoi length associated with the $i$-th node is given by
\begin{equation}
l_{i} = \frac{1} {2} ({{l}}^{i-1} + {{l}}^{i} ).
\end{equation}
We next formulate the discrete strain and curvature from the nodal positions according to DDG theory.
The elongation of $i$-th edge along the meridional direction is straightforward,
\begin{equation}
A^{i}_{11} = \frac { l^{i }} { \bar{l}^{i}} ,
\end{equation}
and the stretching of $i$-th edge along the circumferential is related to the expansion of the circle,
\begin{equation}
A^{i}_{22} = \frac { r^{i} } { \bar{r}^{i} } .
\end{equation}
The coupling between the meridional direction and the circumferential direction for stretching is zero, i.e., 
\begin{equation}
A^{i}_{12} = A^{i}_{21} = 0.
\end{equation}
The bending curvature at a material point $i$ has into $2$ components: the curvature along the meridional direction is identical to the curvature formulation of a planar beam, i.e., is related to the relative positions of the three consecutive nodes,
\begin{equation}
B_{11,i} = \frac { 2 \tan ( {\psi_{i}} / {2} )} { l_{i} },
\end{equation}
where $\psi_{i}$ is the turning angle between $\{\mathbf{x}_{i-1}, \mathbf{x}_{i}, \mathbf{x}_{i+1}\}$ (Fig.~\ref{fig:modelPlot}(d)); 
the one along the circumferential direction is determined by the change of direction of the surface normal, as
\begin{equation}
B_{22,i} = \frac { \cos (\phi_{i}) } { r_{i} },
\end{equation}
where $\phi_{i}$ is the orientation angle, which is evaluated by the surface normal vector, $\mathbf{n}_{i}$, and the $R$ axis, as displayed in Fig.~\ref{fig:modelPlot}(d).
Similarly, the coupling between the meridional direction and the circumferential direction for bending is zero, i.e., 
\begin{equation}
B_{12, i} = B_{21, i} = 0.
\end{equation}

\paragraph{Elastic energy} The total elastic energy of a 1D axisymmetric shell in discrete form is the sum of stretching and bending energies, as 
\begin{equation}
U^{\mathrm{ela}} = \sum_{i=0}^{N-2} \; \pi \; \bar{r}^{i} \; \left( \bm{\epsilon}^{i} \cdot \mathbb{D}_{s} \cdot \bm{\epsilon}^{i} \right) \; \bar{l}^{i} + \sum_{i=1}^{N-2} \; \pi \; \bar{r}_{i} \; \left( \bm{\kappa}_{i} \cdot \mathbb{D}_{b} \cdot \bm{\kappa}_{i} \right) \; \bar{l}_{i} ,
\end{equation}
where the discrete local strain tensors are 
\begin{equation}
\bm{\epsilon}^{i} = 
\begin{bmatrix}
A^{i}_{11} - \bar{A}^{i}_{11} & 0 \\
0 & A^{i}_{22} - \bar{A}^{i}_{22}
\end{bmatrix}
\quad \text{and} \quad
\bm{\kappa}_{i} =
\begin{bmatrix}
B_{11,i} - \bar{B}_{11,i} & 0 \\
0 & B_{22,i} - \bar{B}_{22,i}.
\end{bmatrix}.
\end{equation}
The internal force vector and the associated local stiffness matrix are related to the variation (and the second variation) of the total potential,
\begin{equation}
\mathbf{F}^{\mathrm{ela}} = - \nabla U^{\mathrm{ela}} 
\quad \text{and} \quad
\mathbb{K}^{\mathrm{ela}} = \nabla \nabla U^{\mathrm{ela}} ,
\end{equation}
and we use the chain rule to derive the closed-form expression of the elastic force vector stiffness matrix in terms of the degree of freedom vector $\mathbf{q}$ defined in Eq. \eqref{dofVector}.

\paragraph{Inertial term} The mass matrix of our discrete system is diagonal and time-invariance, i.e., the mass of $i$-th node is,
\begin{equation}
m_{i} = 2 \pi t \; \bar{r}_{i} \; \bar{l}_{i} \; \rho,
\end{equation}
where $\rho$ is the material density.
The time-invariant diagonal mass matrix $\mathbb{M} \in \mathbb{R}^{2N \times 2N}$ can be later implemented.

\paragraph{Equations of motion} Using the force balance and the implicit Euler method to solve the discrete dynamic equations of motion and update the DOF vector from $t= t_{k}$ to $t= t_{k+1}$, we have
\begin{equation}
\mathcal{E} \equiv \mathbb{M} 
 \ddot{\mathbf{q}}(t_{k+1}) - \mathbf{F}^{\text{dam}}(t_{k+1}) - \mathbf{F}^{\text{ela}}(t_{k+1}) - \mathbf{F}^{\text{ext}}(t_{k+1}) = \mathbf{0}
\label{eq:EOM}
\end{equation}
with
\begin{equation}
\begin{aligned}
{\mathbf{q}}(t_{k+1}) &= {\mathbf{q}}(t_{k}) + \dot{\mathbf{q}}(t_{k+1}) \; \delta t \\
\dot{\mathbf{q}}(t_{k+1}) &= \dot{\mathbf{q}}(t_{k}) + \ddot{\mathbf{q}}(t_{k+1}) \; \delta t,
\end{aligned}
\end{equation}
where $ \delta t $ is the time step size, $\mathbf{F}^{\text{ext}}$ is the external force vector, e.g., gravity force, pressure, contact force, and magnetic actuation, and $\mathbf{F}^{\text{dam}}$ is the damping force vector (with damping coefficient defined as $\xi$),
\begin{equation}
\mathbf{F}^{\text{dam}} = - \xi \mathbb{M} \dot{\mathbf{q}}.
\end{equation}
At time step $t_{k+1}$, a new solution is first guessed on the basis of the previous state, i.e.,
\begin{equation}
\mathbf{q}^{(1)}(t_{k+1}) = \mathbf{q}(t_{k}) + \dot{\mathbf{q}}(t_{k}) \; \delta t.
\end{equation}
It is then optimized by utilizing the gradient descent algorithm, such that the new solution at the $(n+1)$-th step is
\begin{equation}
\mathbf{q}^{(n+1)}(t_{k+1}) = \mathbf{q}^{(n)}(t_{k+1}) - ( {\mathbb{M}} / {\delta t^2} + \xi \mathbb{M} / {\delta t} + \mathbb{K}^{\text{ela}} + \mathbb{K}^{\text{ext}}) \backslash \mathcal{E}^{(n)}.
\label{eq:newtonMethod}
\end{equation}
We update the current time step and move forward until the solution is within a prescribed tolerance.
Similar to the 1D rod model, the Jacobian matrix for the 1D axisymmetric shell framework is a banded matrix, which can be solved in linear complexity \cite{huang2023discrete}.
This numerical framework is established for dynamic simulation. However, it can also be used for static equilibrium analysis under arbitrary loading and boundary conditions when damping is introduced into the system as an external dissipative force, and can automatically capture stable equilibrium configurations and avoid unstable equilibrium patterns if perturbations are added into the system \cite{han2003study}.
This framework is well-suited for the analysis of buckling and snapping behaviors \cite{huang2023bifurcations}, which frequently involve multi-equilibria and discontinuous folds.

\section{Model validation}\label{sec:Results}

To validate this 1D axisymmetric shell model, we compare the numerical results from our simulations with existing solutions.
The shell radius is $R$, shell thickness is $h$, Young's modulus is $E$, Poisson ratio is $\nu$, and bending rigidity is $D = Eh^3 / 12 (1 - \nu^2)$.
Damping is incorporated in the dynamical system and the nonlinear equilibrium configuration is derived through the dynamic relaxation method.
The shell contour is discretized along its meridional direction into $N=200$ nodes, the time step size is set to be $ \delta t=1$ms, and the relative tolerance error for the stop is $1e-6$.
The proposed numerical framework is implemented within \textit{C/C++} platform and is optimized by compiling with \textit{Eigen}~\cite{eigenweb}, \textit{BLAS}~\cite{blackford2002updated}, and \textit{LAPACK}~\cite{lapack99};
the sparse linear system of equations is solved by \textit{PARDISO} package from \textit{Intel's oneAPI Math Kernel Library} (MKL)~\citep{krainiuk2021oneapi}.

Two cases for validation are considered: (i) shell buckling under externally applied pressure loading and (ii) shell eversion under external pressure loading.
We also discuss the numerical performance of our discrete model.
The small strain and moderate rotation shell theory adopted for validation is summarized in 
\blue{Appendix A}.

\subsection{Shell buckling under external mechanical pressure loading}

The buckling of a spherical shell with precisely engineered imperfections subjected to a uniform pressure $p$ was first numerically studied by \cite{lee2016geometric,marthelot2017buckling}, followed by the studies of indentation of depressurized spherical shells \cite{pezzulla2019weak,marthelot2017buckling}.
The equivalent force vector on the $i$-th node in the 1D shell model, $\mathbf{x}_{i}$, subjected to a pressure difference between the inner and outer surfaces, $p$, is formulated as
\begin{equation}
\mathbf{F}^{\mathrm{pre}}_{i} = - 2 \pi p \; r_{i} \; l_{i} \; \mathbf{n}_{i}.
\end{equation}
Note that the force vector needs to be updated with configuration to achieve a live pressure force, such that this force is configurational-related and is not conservative.

\begin{figure}[h!]
  \centering
  \includegraphics[width=0.9\textwidth]{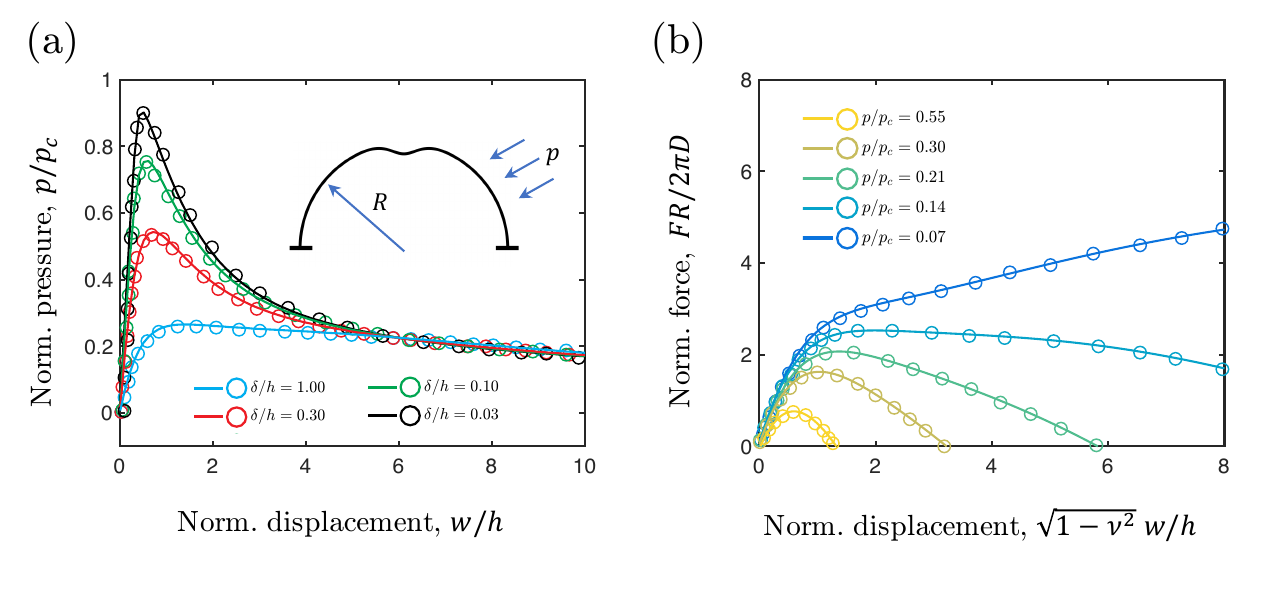}
  \caption{Response of hemispherical shell under external pressure loading. (a) Normalized external pressure, $p/p_c$, as a function of normalized apex displacement, $w/h$, with different imperfections, $\delta / h $. (b) Normalized probe force, $FR/2 \pi D$, as a function of normalized apex displacement, $\sqrt{1-\nu^2} w/h$, under different external pressures, $p/p_{c}$.}
  \label{fig:validateStaPlot}
\end{figure}

Here, we consider the clamped boundary condition at the base and axisymmetry at the pole (illustrated in the inset of Fig.~\ref{fig:validateStaPlot}(a)).
The Dirichlet boundary condition is used to achieve the constraint at the boundary.
The undeformed configuration of a spherical shell with imperfection is given by \cite{lee2016geometric} 
\begin{equation}
\bar{x}(\theta) = \tilde{R} \cos \theta
\quad \text{and} \quad
\bar{z}(\theta) = \tilde{R} \sin \theta,
\end{equation}
where $\theta \in [0, \pi/2] $ and $\tilde{R} $ is the shell radius with imperfection,
\begin{equation}
\tilde{R} = R - \delta e^{ - (\zeta / \tilde{\zeta})^2 }, \; \mathrm{with} \; \zeta = \pi / 2 - \theta.
\end{equation}
Here, $ \tilde{\zeta} $ is the imperfection width and $\delta $ its amplitude at the apex.
To compare with the existing results, we set $ \tilde{\zeta} = 8.83^{\circ}$ and $\delta / h \in \{0.03,0.1,0.3,1 \}~$\cite{pezzulla2019weak,lee2016geometric}.
As the final results are normalized, we use shell radius $R=1.0$ m, thickness $h=0.02$ m, Young's modulus $E = 1.0$ MPa, and Poisson's ratio $\nu=0.5$ for convenience.

In Fig. ~\ref{fig:validateStaPlot}A, we plot the external pressure, $p$, normalized by the critical buckling pressure, $p_{c}$, as a function of the midpoint displacement, $w$, normalized by the shell thickness, $h$.
Here, the critical buckling pressure is for a spherical shell without imperfection and can be derived as \cite{zoelly1915ueber}
\begin{equation}
p_{c}= \frac {2E} {\sqrt{3 (1-\nu^2)} },
\end{equation}
The results from our discrete numerical model are presented as solid lines, while the results from ODEs in literature \cite{pezzulla2019weak,lee2016geometric} are denoted by open circles.
The normalized pressure always increases initially and reaches a maximum (i.e. the buckling load) followed by a monotonic decrease as the displacement increases.
Smaller imperfections (e.g., $\delta/h=0.03$) lead to a sharper initial increase in pressure and higher buckling load.
Excellent agreement is observed between our 1D discrete shell model predictions and the solutions using ODEs, providing strong validation of our model.

In Fig.~\ref{fig:validateStaPlot}(b), we consider the buckling of a pressurized hemispherical shell subjected to a probing force.
Again, for convenience, we choose shell radius $R=1.0$ m, thickness $h=0.02$ m, Young's modulus $E = 1.0$ MPa, and Poisson's ratio $\nu=0.5$.
The initial imperfection is set to be zero to match the existing literature, i.e., $\delta / h=0$.
The pole is constrained to achieve a prescribed displacement, $w$, and the reaction force, $F$, can then be computed based on the statement of equilibrium.
The results of the indentation force $F$ (normalized by $2 \pi h D / R$) versus the normalized additional displacement at the midpoint, $\sqrt{1-\nu^2} w/h$ of an externally pressurized shell under indentation loading are compared with those from \cite{pezzulla2019weak} and \cite{marthelot2017buckling}.
The values of normalized external pressure, $p / p_{c}$, are varied from $0.07$ to $0.55$. 
Under increasing pressure levels, the indentation force versus the displacement curves change from a monotonically increasing to one that exhibits a buckling (maximum) load.
Again excellent agreement is reached between the results of our model (solid lines) and the ODE solutions (open circles). 

\subsection{Shell eversion under mechanical pressure loading}

\begin{figure}[h!]
  \centering
  \includegraphics[width=0.9\textwidth]{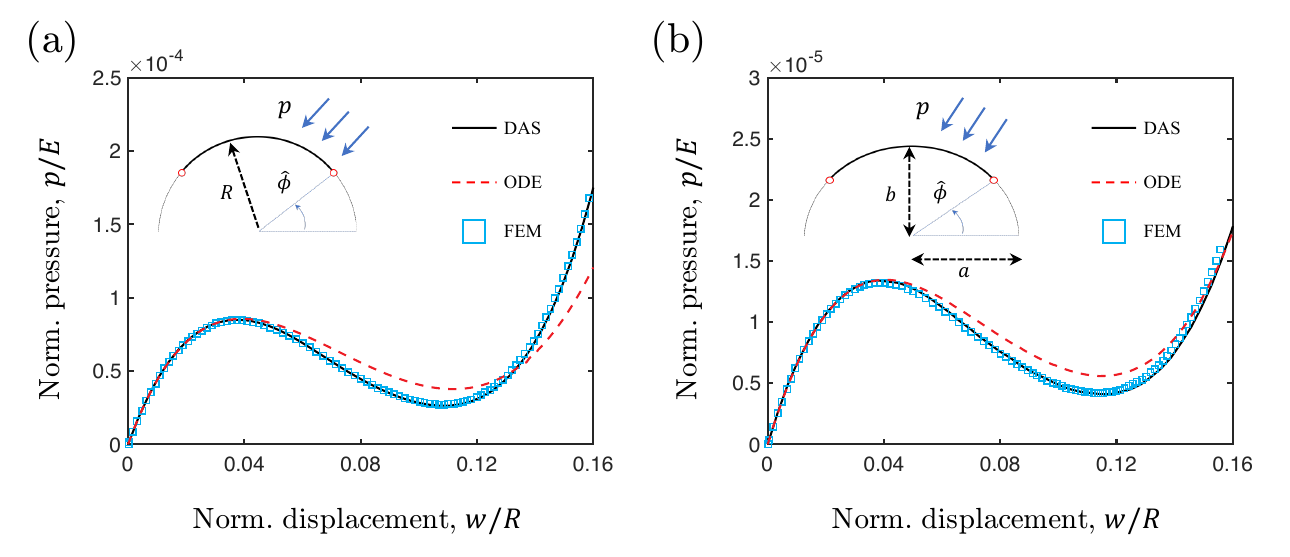}
  \caption{The normalized applied external pressure, $p/E$, as a function of the normalized shell midpoint displacement, $w/R$, compared with the solutions of the classical ODE and FEM simulations for (a) spherical shell and (b) ellipsoidal shell.}
  \label{fig:validateLargePlot}
\end{figure}

We further consider the eversion of both spherical shells and ellipsoidal shells with sliding boundary at the base, in which the rotation of the whole structure is no longer small and the negligence of higher-order terms in ODEs would cause noticeable errors.
We estimate the critical pressure at which eversion occurs.
The normalized pressure, $p/E$, versus the normalized middle-point displacement, $w/R$, is plotted in Fig.~\ref{fig:validateLargePlot}, and compared with both ODEs solutions and FEM simulations \cite{chen2023pseudo}.
Here, the material parameters are Young's modulus $E = 1.0$ MPa and Poisson's ratio $\nu=0.5$, and the geometric parameters of the spherical (Fig.~\ref{fig:validateLargePlot}(a)) and ellipsoidal shells (Fig.~\ref{fig:validateLargePlot}(b)) are selected as $ h/R=0.02$, $\hat{\phi}=3\pi / 8 $ and $ b/a=0.36$, $h/a=0.02$, $\hat{\phi}=17 \pi / 128 $, respectively.
The shell is also assumed to be perfect and there is no imperfection at the pole.

Fig.~\ref{fig:validateLargePlot} shows typical snap-through instability for both spherical and ellipsoidal shells when the structures are deformed to everted configuration, i.e., the pressure first increases to a maximum value, then decreases during the snap-through process and increases again in the everted position.
The maximum pressure corresponds to the load capacity of the shell. For the spherical and ellipsoidal shell configurations considered here, since the minimum load is greater than zero, their deformation is monostable, meaning that they would revert back to their original configurations upon unloading.
The results from our model (black lines) agree well with the FEM simulations for the full range of the displacement, and agree with the ODE solutions before the peak pressure, but deviate beyond the peak, mainly due to the neglect of higher-order terms in ODEs \cite{liu2022buckling,chen2023pseudo}.

\subsection{Numerical performance}

\begin{figure}[t!]
  \centering
  \includegraphics[width=0.9\textwidth]{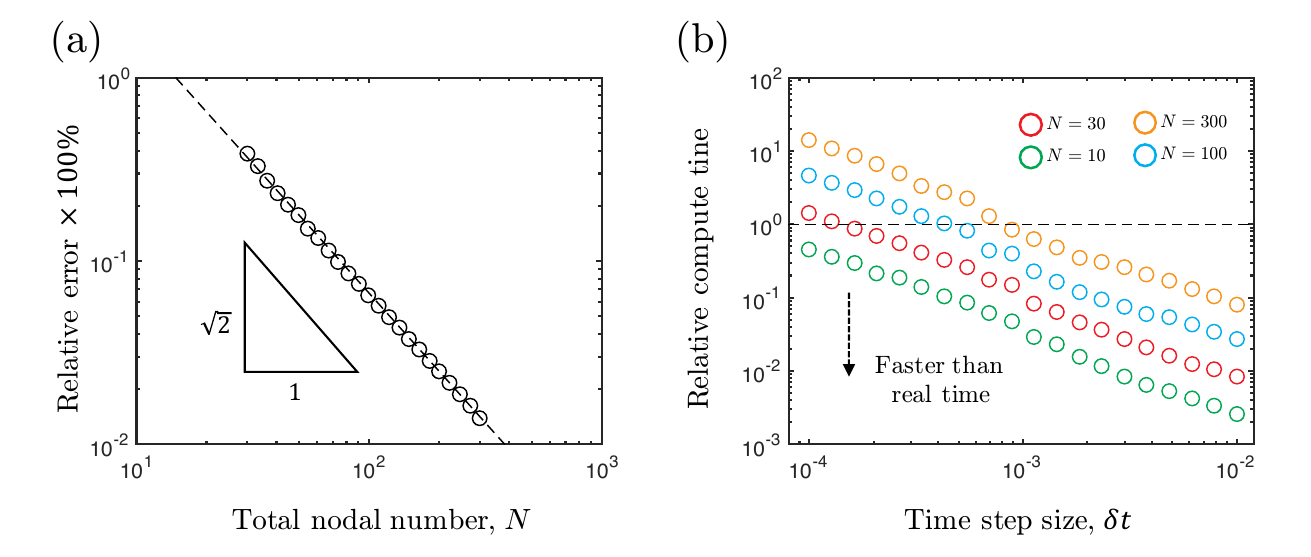}
  \caption{Evaluation of the numerical performance of the discrete 1D shell model. (a) Relative error as a function of nodal number. (b) Computational time over wall-clock time as a function of time step size.}
  \label{fig:conTimePlot}
\end{figure}

Moving forward, we shall evaluate the numerical performance of our 1D shell simulator.
The hemispherical shell eversion under mechanical pressure loading is considered here, which is identical to the scenario shown in Fig.~\ref{fig:validateLargePlot}(a).
We first perform a convergence study of our discrete model.
The critical displacement for snapping is used as a reference.
In Fig.~\ref{fig:conTimePlot}(a), we plot the relative error as a function of the total nodal number, $N$. 
Here, the relative error is defined as 
\begin{equation}
\mathrm{error} =  \frac { || w - w_{\mathrm{real}} ||} {w_{\mathrm{real}}} \times 100\%,
\end{equation}
where $w_{\mathrm{real}}$ is derived when $N=1000$.
The relative error can be less than $1\%$ when $N \ge 30$ and the convergent speed is about $1.414$.
On the other side, in Fig.~\ref{fig:conTimePlot}(b), we show the computational time over wall-clock time, as a function of time step size, $\delta t$, with a different nodal number, $N \in \{10, 30, 100, 300 \}$.
The simulations are performed on a single thread of Intel Core $i7-6600U$ Processor @ $3.4$ GHz.
We can see that our 1D shell framework can be faster than real-time (e.g., when $N=100$ and $\delta t = 1$ ms) on a desktop processor, which is a prerequisite for both digital twins and online controls of soft machines.

\section{Model application}

In this section, we demonstrate our proposed DDG-based numerical method for 1D shell can be used for the analysis of complex engineering problems by incorporating complex loading conditions. 
Two cases are adopted for demonstration: (i) shell snapping and jumping under mechanical contact loading and, (ii) shell buckling and snapping triggered by magnetic actuation.

\subsection{Contact-induced shell snapping and jumping}

Mechanical contact or indentation on shell structures is common in both natural system and engineering applications, e.g., cell diffusion and soft actuators.
We here demonstrate our 1D shell model can be easily incorporated with the boundary nonlinear contact for the investigation of the axisymmetric shell under contact load.

\subsubsection{Numerical Method}

When a node, $\mathbf{x}_{i} $, forms contact with the other surface, a force is generated at the point.
We employ the incremental potential contact method to capture the nonlinear contact boundary between the deformable shell and a rigid surface ~\citep {li2020incremental}.
A smooth log-barrier potential $U^{\mathrm{con}}_{i}$ with $C^2$ continuity is employed when the $i$-th node is within a critical distance with the rigid surface ~\citep{li2020incremental},
\begin{equation}
U^{\mathrm{con}}_{i} =
\begin{cases}
- K_{c} \; ( d_{i} - \hat{d})^2 \; \log( { d_{i} } / {\hat{d}}) \; &\mathrm{when} \; 0 \leq d_{i} < \hat{d}, \\
{0} \; & \mathrm{when} \; d_{i} \geq \hat{d},
\end{cases}
\label{eq:barrierContact0}
\end{equation}
where $d_{i}$ is the minimum distance between $i$-th node and the target object, $\hat{d}$ is the barrier parameter, and $K_{c}$ is the contact stiffness.
The total contact potential is the sum of all contact elements, 
\begin{equation}
U^{\mathrm{con}} = \sum U^{\mathrm{con}}_{i}.
\end{equation}
The contact force is thus the first variation of the proposed potential.
The log-barrier force is zero when the distance is larger than $\hat{d}$. The repulsive interaction gradually increases as the distance decreases within $\hat{d}$. The repulsive force goes to infinite if $ d_{i} $ approaches zero.
The gradient of the contact force vector can be derived analytically due to the $C^2$ continuity of the contact potential, i.e., the force and Jacobian can be formulated in a manner similar to the elastic potentials,
\begin{equation}
\mathbf{F}^{\mathrm{con}} = - \nabla U^{\mathrm{con}} 
\quad \text{and} \quad
\mathbb{K}^{\mathrm{con}} = \nabla \nabla U^{\mathrm{con}} .
\end{equation}
which can be included in the equations of motion in Eq. \eqref{eq:EOM} and Eq. \eqref{eq:newtonMethod}.

\subsubsection{Contact-induced shell snapping}

\begin{figure}[h!]
  \centering
  \includegraphics[width=1.0\textwidth]{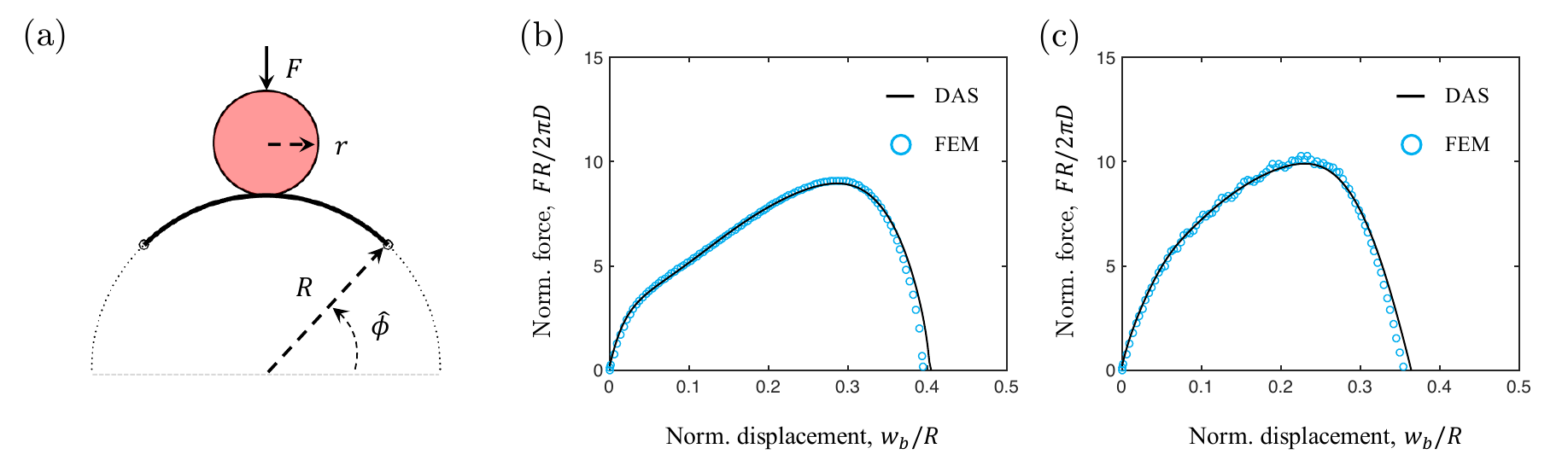}
  \caption{Contact-induced shell snapping. (a) Numerical model setup. The normalized force-displacement curve for (b) $r/R = 0.5$ and (c) $r/R = 1.1$. The curves are derived from our 1D shell model, and the open circles are obtained from FEM simulation.}
  \label{fig:contactShellPlot}
\end{figure}

In this subsection, we use our numerical method to study the shell snapping under a rigid spherical indenter, referring to Fig.~\ref{fig:contactShellPlot}(a).
In these simulations, the meridional angle $\hat{\phi} = 0.8$, shell radius $R = 1.0$ m, shell thickness $h = 0.02$ m, and the relative indenter size $r/R \in \{ 0.5, 1.1\}$ are considered. The material parameters such as Young's modulus $E=1$ MPa and Poisson's ratio $\nu=0.5$ are randomly selected as the problem is purely geometric. The contact interface is assumed to be friction-free, and the contact parameters are $K_{c} = 1e8$ Pa and $\hat{d} = 1e-3$ m after a convergence study, as shown in \blue{Appendix B}.

We manually move the spherical indentor and use continuous contact detection during the numerical simulation, and the sliding boundary at the base is employed to achieve the snap-through mechanics.
The reaction force for the indentor, $F$, can be calculated based on the statement of force equilibrium.
The shape of the shell evolves as the increasing of compressive displacement, $w_b$; when $w_b$ reaches a critical value, the spherical shell would suddenly evert and snap to another equilibrium configuration. 
We observe that the indenter size plays an important role in the shell behavior. When the indenter radius is small (e.g., case with $r/R = 0.5$, Fig.~\ref{fig:contactShellPlot}(b)), the shell pole always remains in contact with the indenter until snapping. However, for larger indenters (case with $r/R = 1.1$, Fig.~\ref{fig:contactShellPlot}(c)), the shell pole detaches from the indenter upon reaching a certain indentation displacement before snapping. This detachment is accompanied by the change of the contact point from the pole towards the outer edge. Detailed dynamic rendering can be found in \blue{Movie-S1} and \blue{Movie-S2}.
For cross-validation, 3D FEM analysis with ABAQUS is also carried out.
Excellent agreement can be found between the FEM simulation and our 1D discrete model.
Details of the FEM simulation can be found in \blue{Appendix C}.

\begin{figure}[t!]
  \centering
  \includegraphics[width=0.65\textwidth]{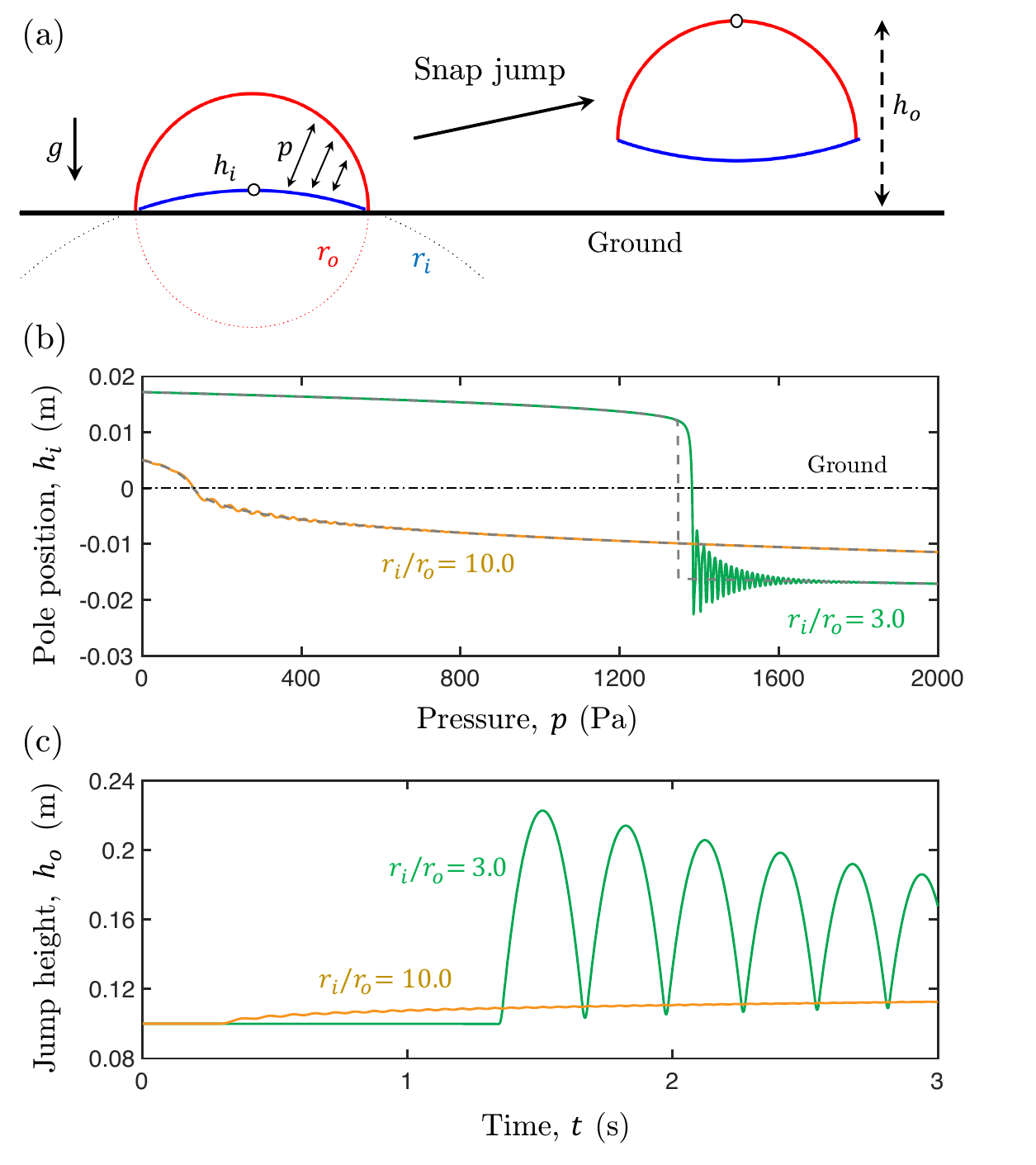}
  \caption{Pneumatic-induced shell jumping. (a) Numerical model setup. (b) The inner shell pole position as a function of the pressure load. The dashed line is from static analysis and the solid line is from the dynamic simulation. Here we ignore the existence of ground. (c) The outer shell pole height as a function of time.}
  \label{fig:jumpShellPlot}
\end{figure}

\subsubsection{Pneumatic-induced shell jumping}

Next, we use a shell-based jumper as an example to demonstrate that our powerful numerical tool is ready for the use of real engineering applications, e.g., soft actuators and soft robots.
Referring to Fig.~\ref{fig:jumpShellPlot}(a), a system comprised of two connected spherical shells is adopted here \cite{gorissen2020inflatable}.
By increasing the pressure between two shells, while the outer shell (i.e., the red one in Fig.~\ref{fig:jumpShellPlot}(a)) maintains its shape, the inner shell (the blue one) may experience a snap-through behavior such that the system can jump up.
To realize the jumping behavior, the contact between the rigid surface and the inner shell needs to be included.
The physical and geometric parameters are selected as follows: outer shell radius $r_{o} = 0.1$ m, inner shell radius $r_{i} \in \{0.3, 1.0\}$ m, shell thickness $h=3$ mm, gravity $g=-10$ m/s$^2$, Young's modulus $E= 50$ MPa, Poisson's ratio $\nu=0.5$, material density $\rho=1000 $ kg/m$^3$, and damping coefficient $\xi = 0.01$.
The contact parameters are identical to the previous study, i.e., $K_{c} = 1e8$ Pa and $\hat{d} = 1e-3$ m.
The meridional range angle of the outer shell is fixed as $\hat{\phi}_{o} = 0.0$, and the meridional angle of the inner shell will be adjusted to its radius, i.e.,
\begin{equation}
\hat{\phi}_{i} = \arcsin (r_{o} / r_{i}).
\end{equation}

We first consider the dynamic snap-through of the inner shell (which serves as actuating unit of the jumper), where the existing ground is ignored.
In Fig.~\ref{fig:jumpShellPlot}(b), we plot the dependence of the pole position, $h_{i}$, on the actuation pressure, $p$.
The loading rate is set as $\dot{p} = 1000$ Pa/s.
When the ratio between the inner shell and outer shell is larger, e.g., $r_{i} / r_{o} = 10$ (with $r_{o} = 0.1$ m fixed), the inner shell is relatively shallow, such that the pressure-height curve (see the yellow line in Fig.~\ref{fig:jumpShellPlot}(b)) is monotonous and no snap-through observed.
On the other side, if the radius ratio is small, e.g., $r_{i} / r_{o} = 3$, the inner shell becomes deeper, and, therefore, it is bistable. As shown by the green line in Fig.~\ref{fig:jumpShellPlot}(b), and a snap-through transition can be observed.
The static continuation is also performed as a reference, where the inertial and damping terms are removed from the Eq. \eqref{eq:EOM}.
Delay bifurcation can be found if the dynamic terms are considered \cite{liu2021delayed}.

Moving forward, by including the contact and collision behaviors between the acting shell and the rigid surface, we can simulate the pneumatic-induced jumping behaviour.
In Fig.~\ref{fig:jumpShellPlot}(c), we plot the maximum height of the system, $h_{o}$, as a function of time, $t$, for both shallower and deeper inner shells.
As expected, the system with a deeper inner shell can jump to a certain height, and reciprocating bounce can be observed, as shown by the green line in Fig.~\ref{fig:jumpShellPlot}(c); while the system with a shallower inner surface can only increase its height smoothly (see the yellow line in Fig.~\ref{fig:jumpShellPlot}(c)).
The dynamic renderings for these two scenarios are available in \blue{Movie-S3} and \blue{Movie-S4}.

\subsection{Magnetic-induced shell buckling and snapping}

If the shell is made of a hard-magnetic soft material, it can be actuated by a magnetic field \cite{lucarini2022recent}.
Unlike the mechanical loadings, magnetic actuation exhibits interesting features, e.g., quick response, remote controllability, and reversible deformation \citep{huang2023discrete,lucarini2022recent}, and has been widely adopted in soft robots \citep{huang2023modeling,wang2020hard} and soft machines \cite{abbasi2023leveraging, chen2024magnet}.
We here demonstrate our 1D shell model can be easily extended to incorporate the magneto-elastic constitutive law for the investigation of the axisymmetric deformation of magnetic shell structure.

\subsubsection{Numerical Method}

The magnetization of the $i$-th edge element is denoted by $\mathbf{B}^{i}_{r} \in \mathbb{R}^{2 \times 1}$, and the external magnetic field applied to this element is $\mathbf{B}^{i}_{a} \in \mathbb{R}^{2 \times 1}$.
From the Helmholtz free energy theory, the total magneto potential for a thin shell is
\begin{equation}
 U^{\mathrm{mag}} = - \frac{2 \pi t}{\mu_{0}} \sum_{i = 0}^{N-2} \bar{r}^{i} \; \bar{l}^{i} \; \left( \mathbb{B}^{i} : \mathbb{E}^{i} + \mathbb{B}^{i} : \mathbb{N}^{i} \right) ,
 \label{magPotential}
\end{equation}
where $\mu_{0}$ is the vacuum permeability, and
\begin{equation}
\begin{aligned}
\mathbb{B}^{i} &= \mathbf{B}^{i}_{r} \otimes \mathbf{B}^{i}_{a} \\
\mathbb{E}^{i} &= \frac{{l}_{i}} {\bar{l}_{i}} \left( \mathbf{e}^{i} \otimes \bar{\mathbf{e}}^{i} \right) \\
\mathbb{N}^{i} &= \mathbf{n}^{i} \otimes \bar{\mathbf{n}}^{i}.
\end{aligned}
\end{equation}
Here, $\otimes$ denotes the tensor product.
The external magnetic force and the associated Hession matrix can be derived as
\begin{equation}
\mathbf{F}^{\mathrm{mag}} = - \nabla U^{\mathrm{m}} 
\quad \text{and} \quad
\mathbb{K}^{\mathrm{mag}} = \nabla \nabla U^{\mathrm{m}}.
\end{equation}
which are later used in solving the equations of motion in Eq. \eqref{eq:EOM} and Eq. \eqref{eq:newtonMethod}.
It is worth noting that here we use the traditional $F$-based formulation for magnetic potential, and the novel $R$-based method can be easily achieved by using $ \mathbb{T}^{i} $ instead of $ \mathbb{E}^{i} $ in Eq.~(\ref{magPotential}) \cite{yan2023reduced},
\begin{equation}
\mathbb{T}^{i} = \mathbf{t}^{i} \otimes \bar{\mathbf{t}}^{i},
\end{equation}
in which the volume change of the shell element is erased and only rotation is taken into account.

\subsubsection{Magnetic-induced shell buckling}

\begin{figure}[h!]
  \centering
  \includegraphics[width=1.0\textwidth]{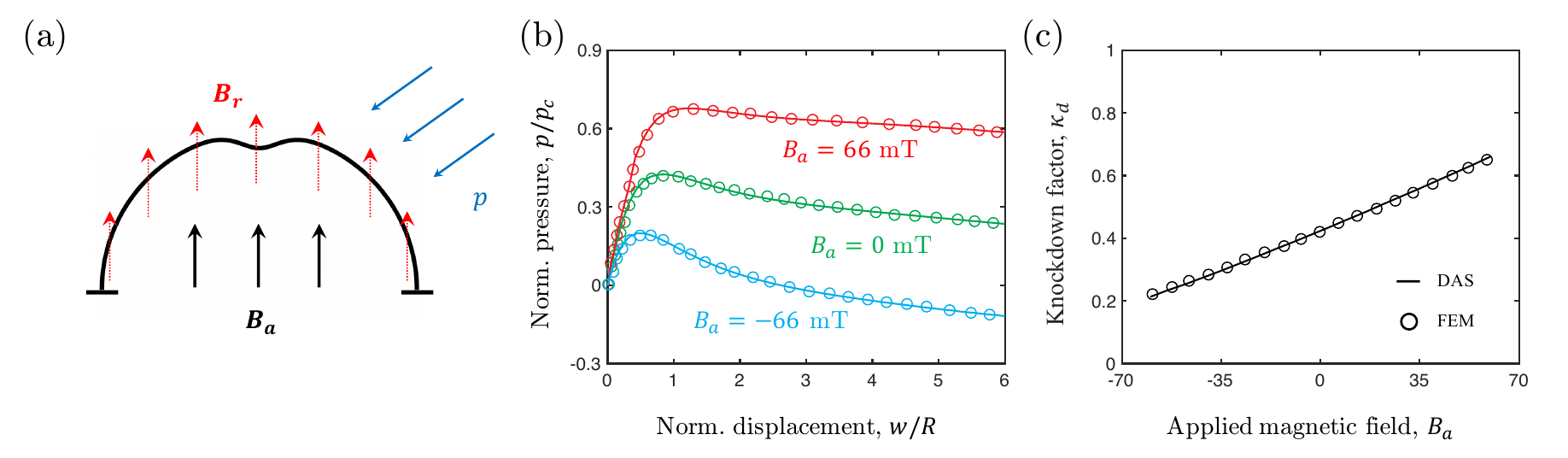}
  \caption{Magnetic-induced shell buckling. (a) Numerical model setup. (b) Normalized pressure as a function of normalized pole displacement. (c) Relationship between the knockdown factor and the external magnetic strength. The curves are from our 1D shell model, and the open circles are from FEM simulation adopted from \cite{yan2021magneto}.}
  \label{fig:magShellPlot}
\end{figure}

We first consider the shell buckling problem under a combined mechanical and magnetic load \cite{yan2021magneto}.
The physical and geometric parameters are adopted from the literature \cite{yan2021magneto}: shell radius $R=25.4$ mm, radius-to-thickness ratios ratio $R/h = 91.3$, Young's mdlus $E = 1.69$ MPa, Poisson's ratio $\nu=0.443$, permanent magnetization $B_{r} = 33$ mT, external magnetic field $B_{a} \in[-66, 66] $ mT, and permeability of vacuum $\mu_{0} = 4 \pi \times 10^{-7}$ H/m.
Referring to Fig.~\ref{fig:magShellPlot}(a), both ${B}_{a}$ and ${B}_{r}$ are along the vertical ($z$) axis, and ${B}_{a}$ (and ${B}_{r}$) is the $\mathcal{L}_{2}$ norm of the $\mathbf{B}_{a}$ (and $\mathbf{B}_{r}$).
To compare with existing results \cite{yan2021magneto}, the imperfect shell contour radius is slightly modified from the previous case, given by
\begin{equation}
\tilde{R} = 
\begin{cases}
R - \delta (1 - {\zeta^2} / {\tilde{\zeta}^2} )^2 \; &\mathrm{when} \; \zeta \in [0, \tilde{\zeta} ], \\
R \; & \mathrm{when} \; \zeta \in (\tilde{\zeta}, \pi/2),
\end{cases}
\label{eq:barrierContact}
\end{equation}
where $\tilde{\zeta}$ is the defect width and $\delta$ is the imperfection size.
We use $\tilde{\zeta} = 11.8^{\circ}$ and $\delta/h=0.39$ to match the existing literature \cite{yan2021magneto}.
Also, the clamped boundary condition at the base and axisymmetry condition at the pole are used.

The buckling of hard-magnetic elastic shells under combined external pressure and magnetic field loading is examined.
In Fig.~\ref{fig:magShellPlot}(b), we plot the relationship between the normalized pressure, $p/p_c$, as a function of the normalized pole displacement, $w/R$, with a fixed magnetic field, $B_{a} \in \{-66, 0, 66 \}$ mT.
When the direction of the shell magnetization, $\mathbf{B}_{r}$, is identical to the direction of the external magnetic field, $\mathbf{B}_{a}$, the buckling strength of the shell can be enlarged, i.e., the maximum pressure for buckling instability is larger.
On the other side, if the direction of the shell magnetization, $\mathbf{B}_{r}$, is opposite to the direction of the external magnetic field, $\mathbf{B}_{a}$, the buckling load would decrease. 
Next, in Fig.~\ref{fig:magShellPlot}(c), we plot the knockdown factor, $ \kappa_{d} = p_{\mathrm{max}}/p_{c}$, as a function of the applied magnetic field, $B_{a}$.
Depending on whether the direction of the applied magnetic field $\mathbf{B}_{a}$ is the same as or opposite to the direction of the shell magnetization $\mathbf{B}_{r}$, the buckling pressure of hard-magnetic shells can be higher or lower than that without the magnetic field. 
The knockdown factor $\kappa_{d}$ exhibits approximately linear dependence on the applied flux density $B_a$.
The results from our model (the solid line) agree well with the predictions from FEM simulation (open circles) \cite{yan2021magneto}, validating our approach in predicting shell buckling under magnetic actuation.

\subsubsection{Magnetic-induced shell snapping}

\begin{figure}[h!]
  \centering
  \includegraphics[width=1.0\textwidth]{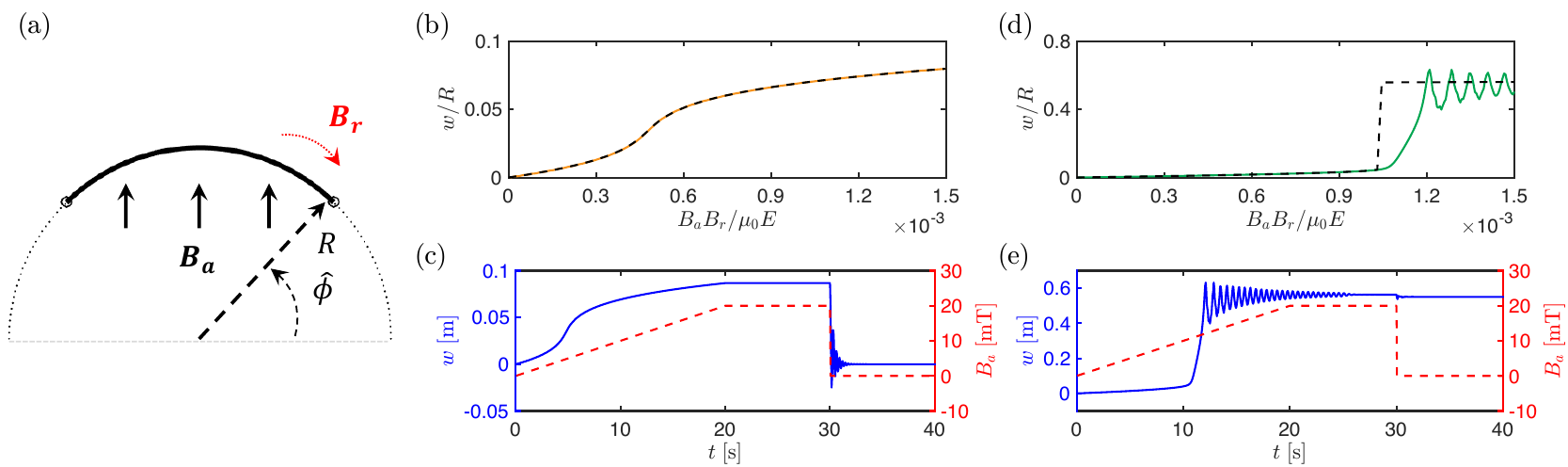}
  \caption{Magnetic-induced shell snapping. (a) Numerical model setup. (b) Normalized pole displacement as a function of normalized magnetic strength for a shallow shell, $\hat{\phi}=1.3$. (c) Pole displacement as well as the external magnetic field as a function of time for a shallow shell, $\hat{\phi}=1.3$. (d) Normalized pole displacement as a function of normalized magnetic strength for a deep shell, $\hat{\phi}=0.8$. (e) Pole displacement as well as the external magnetic field as a function of time for a deep shell, $\hat{\phi}=0.8$.}
  \label{fig:magSnapPlot}
\end{figure}

Next, we turn our attention to the shell snapping problem under magnetic actuation \cite{stewart2023magneto}.
Here, we consider the shell with radius $R=1.0$ m, thickness $h = 0.02$ m, meridional angle $\hat{\phi} \in \{0.8, 1.3\}$, Young's modulus $E = 1.0$ MPa, Poisson's ratio $\nu=0.5$, material density $\rho = 1000$ kg/m$^3$, damping ceofecinet $\xi=0.1$, permanent magnetization $B_{r} = 100$ mT, and permeability of vacuum $\mu_{0} = 1e-6$ H/m. The applied external magnetic field is time-dependent, $B_{a} (t)$, which is defined as
\begin{equation}
B_{a}(t) = 
\begin{cases}
\dot{B}_{a} \cdot t \; \mathrm{mT} \; &\mathrm{when} \; 0 \; \mathrm{ s} \leq t < 20.0 \; \mathrm{ s}\\
20.0 \; \mathrm{mT} \; &\mathrm{when} \; 20.0 \; \mathrm{ s} \leq t < 30.0 \; \mathrm{ s} \\
0.0 \; \mathrm{mT} \; &\mathrm{when} \; 30 \; \mathrm{ s} \leq t < 40 \; \mathrm{ s}, 
\end{cases}
\end{equation}
where $\dot{B}_{a} = 1.0$ mT/s is the loading rate.
The directions of the permanent magnetization and the external magnetic field are shown in Fig.~\ref{fig:magSnapPlot}(a), i.e., the permanent magnetization of the shell $\mathbf{B}_{r}$ is assumed to be along the surface meridional direction, i.e., the permanent magnetization of $i$-th edge element is 
\begin{equation}
\mathbf{B}_{r}^{i} = - {B}_{r} \mathbf{t}^{i},
\end{equation}
which is easier for manufacturing.
The external magnetic field $\mathbf{B}_{a}$ is applied along the vertical ($z$) axis.
Also, the shell is assumed to be perfect (i.e., there is no imperfection at the pole).
Similarly, to achieve the snap-through transition, the sliding boundary at the base is selected.

In Fig.~\ref{fig:magSnapPlot}(b), we plot the dependence of the normalized pole displacement, $w/R$, on the normalized magnetic strength, $B_a B_r / \mu_0 E$, for a relatively shallow shell, $\hat{\phi} = 1.3$.
The black dashed line is obtained from static analysis, and the yellow solid line is from the dynamic simulation.
We also show the load process in Fig.~\ref{fig:magSnapPlot}(c), i.e., the applied magnetic field, $B_a$, as a function of time, $t$, see the red dash line. 
The corresponding relationship between the pole displacement, $w$, and the time, $t$, is also given by blue solid line.
As we can see from Fig.~\ref{fig:magSnapPlot}(b), the load-displacement curve is monotonous and no snap-through has been observed, and the shell would go back to the initial configuration when the external magnetic field is canceled out when $t > 30$ s (as indicated by the blue curve in Fig.~\ref{fig:magSnapPlot}(c)).
On the other side, in Fig.~\ref{fig:magSnapPlot}(d), we plot the dependence of the pole displacement $w/R$ on the magnetic strength $B_a B_r / \mu_0 E$ for a relatively shallow shell (i.e., $\hat{\phi} = 0.8$).
The corresponding load process is also given in Fig.~\ref{fig:magSnapPlot}(e), i.e., the relation between $B_a$ and $t$ for the red dash line and the relation between $w$ and $t$ for the blue solid line.
We can see the abrupt change in the load-displacement curve (Fig.~\ref{fig:magSnapPlot}(d)), which is corresponding to the snap-through or shell eversion, and, the shell would remain in the everted configuration even when the external magnetic actuation is deleted after $t=30$ s (see the blue curve in Fig.~\ref{fig:magSnapPlot}(e)).
Some oscillations can be observed during the loading/unloading processes as the dynamic effects (e.g., inertial and damping) have been included in our numerical simulation.
The dynamic rendering for these two cases can be found in \blue{Movie-S5} and \blue{Movie-S6}.

\section{Concluding remarks}

This work presents a 1D shell model based on the discrete differential geometry (DDG) to carry out geometrically nonlinear mechanics analysis of thin elastic shells under axisymmetric constraints.
This DDG-based model enables us to establish a dynamic framework that captures critical bifurcation points and discontinuous snapping phenomena, reaching stable equilibrium configurations under various boundary and loading conditions.
Examples of using the developed model to solve shell problems with known solutions demonstrate its fidelity compared to other methods.
More importantly, when a shell undergoes large deflection and finite rotation, such as eversion, our model shows a clear advantage over classical ODEs in terms of accuracy when validated by FEM simulations, as our numerical model has no restrictions on the deflections or rotations of the mid-surface of the shell.
It was also worth noticing that this method is highly computationally efficient, and can easily incorporate geometrically nonlinear deformations together with boundary nonlinear contacts as well as multi-physical actuation. 
Several examples were later used to demonstrate the effectiveness of our proposed method when handling complex loading conditions.
The accuracy, simplicity, and efficiency of our 1D DDG-based shell model make it an attractive alternative to conventional numerical methods such as FEM for researchers who seek to study nonlinear mechanics of axisymmetric shells.
The model has the potential for various applications in engineering fields and biomechanical phenomena. It paves the way for future research and exploration in mechanical analysis and design of shell structures in both natural environments and engineered systems including, but not limited to, phenomena such as cell fusion, robotic actuating, and submarine collapsing.

\appendix

\section{Classic shell theory}
\label{sec:shellODE}

Here, we review the ODEs for shells using small strain, moderate rotation shell theory \cite{hutchinson2016buckling}.
The position vector $\mathbf{x}$ of a material point with a coordinate $(\theta,\omega)$ on the mid-surface of the undeformed shell can be expressed in the three-dimensional Euclidean space as
\begin{equation}
\mathbf{x}\left(\theta,\ \omega\right)=\left[R\left(\theta\right)\cos{\theta}\cos{\omega}\right]\mathbf{e}_1+\left[R\left(\theta\right)\cos{\theta}\sin{\omega}\right]\mathbf{e}_2+\left[R\left(\theta\right)\sin{\theta}\right]\mathbf{e}_3,
\end{equation}
where $\left\{\mathbf{e}_1,\mathbf{e}_2,\mathbf{e}_3\right\}$ is a group of orthonormal bases in the Euclidean space, $\theta$ is the meridional angle ranging from $\theta_{\mathrm{min}}$ to ${\pi}/{2}$ at the pole, and $\omega$ is the circumferential angle. The displacement of this material point can be written as
\begin{equation}
\mathbf{\delta}\left(\theta,\ \omega\right)=u^\beta\mathbf{x}_{,\beta}+w\mathbf{N},
\end{equation}
where $\mathbf{x}_{,\beta}={\partial\mathbf{x}}/{\partial\beta}$ and $\mathbf{N}$ denote the covariant bases and the normal vector of the mid-surface at the undeformed state, respectively, and $\left(u^\beta,w\right)$ are the corresponding displacements. A Greek index takes on values of $\theta$ and $\omega$, and a repeated Greek index means summation over $\theta$ and $\omega$. Here we only consider axisymmetric deformations. As a result, $u^\theta$ and $w$ are only functions of $\theta$, i.e., $u^\omega=0$, and the rotation about $\mathbf{x}_{,\theta}$, $\varphi_\omega=\varphi^\omega=0$. The corresponding non-zero mid-surface strains and curvature strains under axisymmetric deformation are
\begin{equation}
\begin{aligned}
E_\omega^\omega=u^\theta\Gamma_{\theta\omega}^\omega+b_\omega^\omega w, \\
E_\theta^\theta={u^\theta}^\prime+u^\theta\Gamma_{\theta\theta}^\theta+b_\theta^\theta w+\frac{1}{2}\varphi^2g_{\theta\theta}, \\
K_\omega^\omega=\varphi\Gamma_{\theta\omega}^\omega,\\
K_\theta^\theta=\varphi^\prime+\varphi\Gamma_{\theta\theta}^\theta,
\end{aligned}
\end{equation}
where $\left( \right)^\prime$ denotes ${d\left( \right)}/{d\theta}$, $\varphi=\varphi^\theta=g^{\theta\theta}\varphi_\theta$, $\varphi_\theta$ denotes the rotation about $\mathbf{x}_{,\omega}$,
\begin{equation}
\varphi=-w^\prime g^{\theta\theta}+b_\theta^\theta u^\theta,
\end{equation}
$\Gamma_{\theta\omega}^\omega$ and $\Gamma_{\theta\theta}^\theta$ are Christoffel symbols, $g_{\alpha\beta}$ and $g^{\alpha\beta}$ are the covariant and contravariant components of the first fundamental form of the mid-surface, and $b_\alpha^\beta$ are the mixed components of the second fundamental form of the mid-surface. The mid-surface and curvature strains can be expressed in terms of $u^\theta$ and $w$, or equivalently in terms of $\varphi$ and $w$. Here we choose $\varphi$ and $w$ as the two independent variables by replacing $u^\theta$ with a function of $\varphi$ and $w$. The strain of the shell at an arbitrary position can be expressed as $\varepsilon_\alpha^\beta=E_\alpha^\beta+zK_\alpha^\beta$, where $z$ is the coordinate in the thickness direction of the shell and measured from the mid-surface.

Next, the principle of virtual work is used to derive the equilibrium equations. Let $\delta u_\theta$ and $\delta w$ be the virtual displacements of the mid-surface of the shell. The associated virtual strains can be expressed as $\delta\varepsilon_{\alpha\beta}=\delta E_{\alpha\beta}+z\delta K_{\alpha\beta}$. The internal virtual work (IVW) of the shell is
\begin{equation}
\mathrm{IVW}=\int_{S}\left[N^{\alpha\beta}\delta E_{\alpha\beta}+M^{\alpha\beta}\delta K_{\alpha\beta}\right]dS,
\end{equation}
where $S$ represents the area of the mid-surface of the shell, and the resultant membrane stresses $N_{\alpha\beta}$ and the bending moments $M_{\alpha\beta}$ are
\begin{equation}
\begin{aligned}
N^{\alpha\beta}=\frac{E_0h}{1-\nu^2}\left[\left(1-\upsilon\right)E^{\alpha\beta}+\upsilon E_\gamma^\gamma g^{\alpha\beta}\right], \\
M^{\alpha\beta}=\frac{E_0h^3}{12\left(1-\nu^2\right)}\left[\left(1-\upsilon\right)K^{\alpha\beta}+\upsilon K_\gamma^\gamma g^{\alpha\beta}\right],
\end{aligned}
\end{equation}
where $h$ denotes the thickness of the shell, $E_0$ denotes Young’s modulus (which is $E$ in our main text), and $\upsilon$ is the Poisson’s ratio.
The external virtual work (EVW) due to a uniform live pressure $\Delta P$ acting on the shell is
\begin{equation}
\mathrm{EVW}=\int_{S}\left[\Delta P\varphi\delta u_\theta+\Delta P\left(1+u_{,\gamma}^\gamma+b_\gamma^\gamma w\right)\delta w\right]dS+\oint_{C}\left(T^\theta\delta u_\theta+Q\delta w-M_n{\delta w}_{,n}\right)ds,
\end{equation}
where $T^\theta$ represents the edge resultant traction along $\mathbf{x}_{,\theta}$, $Q$ represents the normal edge force, and $M_n=M^{\alpha\beta}n_\alpha n_\beta$ is the component of the edge moment, $n_\beta$ denotes the components of the unit vector normal to the boundary $C$ tangent to the shell, and $s$ is the length of the edge of the shell. Enforcing IVW = EVW yields the following equilibrium equations
\begin{equation}
\begin{aligned}
-M_{,\alpha\beta}^{\alpha\beta}+N^{\alpha\beta}b_{\alpha\beta}+\left(N^{\alpha\beta}\varphi_\alpha\right)_{,\beta}=\Delta P\left(1+u_{,\gamma}^\gamma+b_\gamma^\gamma w\right), \\
-N_{,\beta}^{\theta\beta}-M_{,\beta}^{\alpha\beta}b_\alpha^\theta-\frac{1}{2}\left(M^{\alpha\beta}b_\alpha^\theta-M^{\theta\alpha}b_\alpha^\beta\right)_{,\beta}+N^{\alpha\beta}\varphi_\alpha b_\beta^\theta=\Delta P\varphi,
\end{aligned}
\end{equation}
where $\left(\right)_{,\alpha}$ and $\left(\right)_{,\alpha\beta}$ are the first and second-order covariant derivatives of $\left(\right)$. Here, $\varphi$ and $w$ are the two independent variables, and the highest order terms are $\varphi^{\prime\prime\prime}$ and $w^{\prime\prime\prime}$, yielding a system of six-order nonlinear ordinary differential equations (ODEs). In order to limit the deformation within the assumption of small strain and moderate rotation for a deep shell when fully everted, we choose the sliding boundary for the shell, i.e., on the boundary at $\theta=\theta_{\mathrm{min}}$, the shell is allowed to slide freely along $\mathbf{e}_1$, but not along $\mathbf{e}_3$. As a result, the traction along $\mathbf{e}_1$ is zero
\begin{equation}
\left(T^\theta\mathbf{x}_{,\theta}+Q\mathbf{N}\right) \cdot \mathbf{e}_1=T^\theta\left(R^\prime\cos{\theta}-R\sin{\theta}\right)+Q\frac{R^\prime\sin{\theta}+R\cos{\theta}}{\sqrt{{R^\prime}^2+R^2}}=0,
\end{equation}
and the displacement along $\mathbf{e}_3$ is zero,
\begin{equation}
\mathbf{\delta} \cdot \mathbf{e}_3=u^\theta\left(R^\prime\sin{\theta}+R\cos{\theta}\right)-\frac{w\left(R^\prime\cos{\theta}-R\sin{\theta}\right)}{\sqrt{{R^\prime}^2+R^2}}=0,
\end{equation}
In addition, the assumption of axisymmetric deformation requires $w^\prime=\varphi=\varphi^{\prime\prime}=0$ at the pole $(\theta={\pi}/{2})$.

Here we solve the ODE set by prescribing the displacement at the pole $w(\theta=\pi/2)$ as the load parameter. With this loading type, the pressure is treated as an extra variable. Correspondingly, an additional ODE, $\Delta P^\prime=0$, is added to the ODE set.

\begin{figure}[h!]
  \centering
  \includegraphics[width=0.9\textwidth]{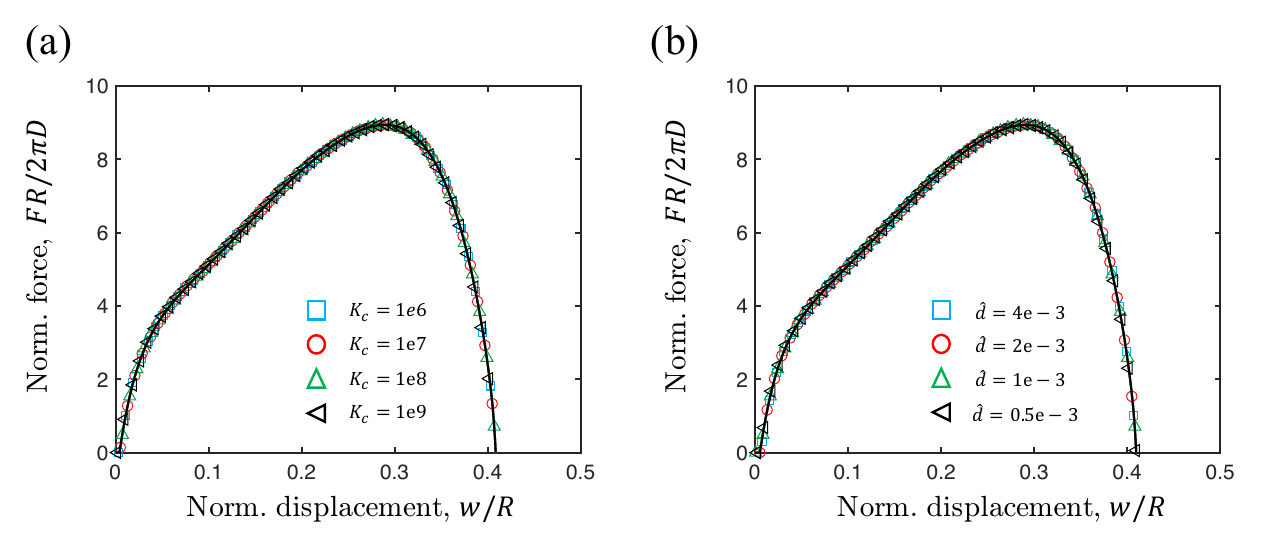}
  \caption{Convergence study for parameters used in the contact model: (a) contact stiffness, $K_{c}$, and (b) threshold distance, $\hat{d}$.}
  \label{fig:convergentPlot}
\end{figure}

\section{Convergence study for contact parameter}
\label{sec:Convergentforcontact}

In this appendix, we perform a convergence study on the numerical parameters used in our incremental potential-based contact model.
There are two parameters, contact stiffness, $K_{c}$, and clamped distance, $\hat{d}$, that are used to achieve the non-penetration condition between the flexible shell and the rigid surface.
Here, we use an axisymmetric shell with $ h/R = 0.02$ and $\hat{\phi} = 0.8$, and the indenter radius is set to be $r/R = 0.5$.
Fig.~\ref{fig:convergentPlot}(a), we fixed the clamped distance as $\hat{d} = 1e-3$ m, and vary the contact stiffness, $K_{c} \in \{1e6, 1e7, 1e8, 1e9\}$ Pa.
All curves can be overlapped with each other, which indicates our numerical results are independent with $K_{c}$.
On the other side, in Fig.~\ref{fig:convergentPlot}(b), the contact stiffness is fixed as $K_{c} = 1e8$, and we change the clamped distance, $\hat{d} \in \{4e-3, 2e-3, 1e-3, 0.5e-3\}$ m, to check the variations.
Again, all identical numerical results prove that our simulations show good convergence on the contact-clamped distance parameters.
In all our numerical experiments, we use $K_{c} = 1e8$ Pa and $\hat{d} = 1e-3$ m.

\section{Finite element analysis}
\label{sec:FEMSimulation}

\begin{figure}[h!]
  \centering
  \includegraphics[width=1.0\textwidth]{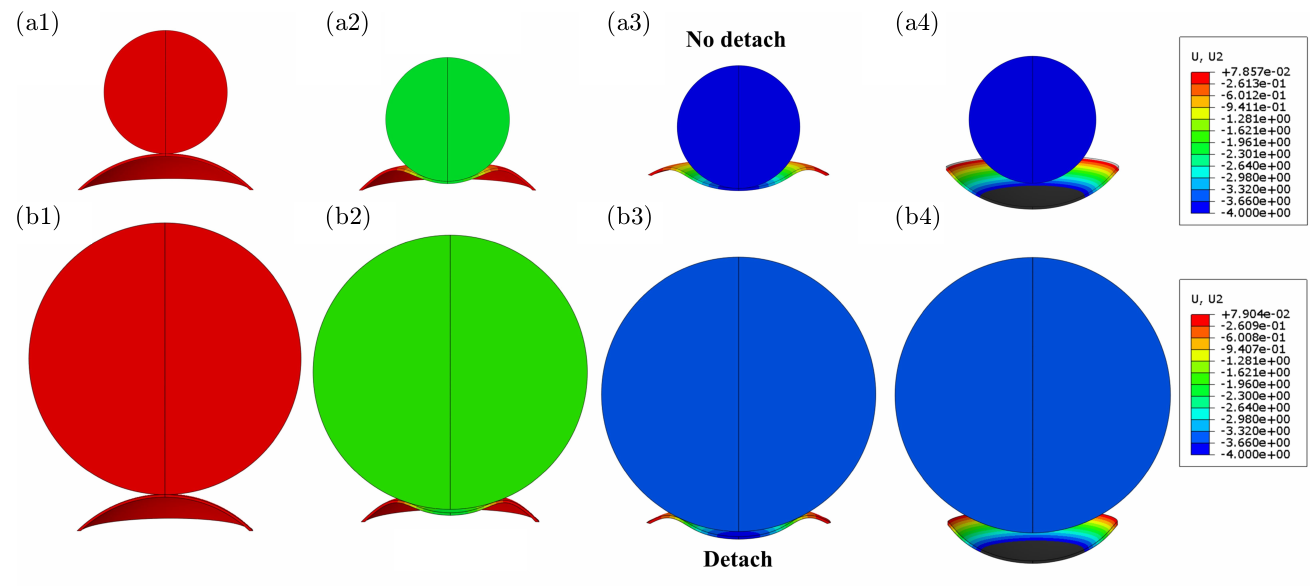}
  \caption{Configuration evaluations of spherical shells under indenter probe actuation. (a) $r/R=0.5$ and (b) $r/R=1.1$.}
  \label{fig:femConfPlot}
\end{figure}

The FEM simulation is performed to investigate the deformation mechanisms of spherical shells under the compression of indenter.
The obtained results are compared with the discrete numerical model to verify the accuracy of the presented model.
Here in the case of indentation loading, the base of the shell is free to flip and slide along the horizontal plane.
Hence, the movement of the shell is restrained vertically by applying a zero vertical displacement ($U2=0$), but other degrees of freedom are free to evolve.
The dynamic responses of the shell under the indentation loading process are simulated using the dynamic implicit solver of Abaqus software.
In order to compute a wide range of strain and deformation of axisymmetric shell, the continuum full integration axisymmetric elements (CAX4) are adopted, which could overcome the hourglassing and severe distortion effect.
The nonlinear deformation such as large deflection and large rotation is also activated.
The material parameters of the shell used in the FEM simulation include elastic modulus $E=20$MPa, and Poisson’s ratio $\nu=0.49$, which would be normalized later for comparison with the results of our discrete numerical model.
The geometric parameters of the indenter and shell are selected as $r/R \in \{0.5, 1.1\}$.
In addition, the model with a frictionless surface is considered in the current study. 
The snapshots from our FEM can be found in  Fig.~\ref{fig:femConfPlot}.

\section{Videos}
\label{sec:video}
We also provide several videos as supplementary materials.

\section*{Acknowledgments}

T. L. acknowledges the support from China Postdoctoral Science Foundation (2022M720721). M.L. acknowledges the Presidential Postdoctoral Fellowship from Nanyang Technological University, Singapore, and the start-up funding from the University of Birmingham. K.J.H. acknowledges the financial supports from Nanyang Technological University, Singapore (Grant M4082428) and Ministry of Education, Singapore under its Academic Research Fund Tier 2 (T2EP50122-0005). We are grateful to Dominic Vella, Zhaohe Dai, and Yingchao Zhang for the useful discussions.

\section*{Conflict of interest}

The authors declare that they have no conflict of interest.

\section*{Availability of data}

The datasets generated during the current study are available from the corresponding author upon reasonable request.

\section*{Availability of Code}
The code generated during the current study is available from the corresponding author on reasonable request.

\bibliographystyle{elsarticle-num}
\bibliography{aShell}

%% Authors are advised to submit their bibtex database files. They are
%% requested to list a bibtex style file in the manuscript if they do
%% not want to use elsarticle-harv.bst.

%% References without bibTeX database:

% \begin{thebibliography}{00}

%% \bibitem must have one of the following forms:
%%   \bibitem[Jones et al.(1990)]{key}...
%%   \bibitem[Jones et al.(1990)Jones, Baker, and Williams]{key}...
%%   \bibitem[Jones et al., 1990]{key}...
%%   \bibitem[\protect\citeauthoryear{Jones, Baker, and Williams}{Jones
%%       et al.}{1990}]{key}...
%%   \bibitem[\protect\citeauthoryear{Jones et al.}{1990}]{key}...
%%   \bibitem[\protect\astroncite{Jones et al.}{1990}]{key}...
%%   \bibitem[\protect\citename{Jones et al., }1990]{key}...
%%   \harvarditem[Jones et al.]{Jones, Baker, and Williams}{1990}{key}...
%%

% \bibitem[ ()]{}

% \end{thebibliography}

\end{document}